\documentclass[journal]{IEEEtranTIE}
\usepackage{graphicx}
\usepackage{cite}
\usepackage{picinpar}
\usepackage{amsmath}
\usepackage{url}
\usepackage{flushend}
\usepackage[latin1]{inputenc}
\usepackage{colortbl}
\usepackage{soul}
\usepackage{multirow}
\usepackage{pifont}
\usepackage{color}
\usepackage{alltt}
\usepackage[hidelinks]{hyperref}
\usepackage{enumerate}
\usepackage{siunitx}
\usepackage{breakurl}
\usepackage{epstopdf}
\usepackage{pbox}

\usepackage[svgnames]{xcolor}

\begin{document}
\title{	A precise detection method for transient micro short-circuit faults of lithium-ion batteries through signal processing }

\author{
	\noindent  
        Hongyu Zhao, Yangyang Xu, Chenglin Liao 
            \\ Email:\href{mailto:Zhaohongyu21@mail.iee.ac.cn}{  
            \color{Blue} 
            \underline{Zhaohongyu21@mail.iee.ac.cn}  
        } \\
        Institute of Electrical Engineering, Chinese Academy of Sciences, 100190  

	}

\maketitle
	
\begin{abstract}
A specific failure mode designated as transient micro-short circuit (TMSC) has been identified in practical battery systems, exhibiting subtle and latent characteristics with measurable voltage deviations. To further improve the safe use of lithium-ion batteries (LIBs), this letter introduces a novel method for the precise detection of this TMSC faults within LIBs. The method applies the continuous wavelet transform (CWT) to voltage and current signals, followed by the identification of micro-scale anomalies through the analysis of the coherence in the wavelet spectrum at specific frequency. Through designed fault experiments, the effectiveness of this method has been verified. Result demonstrates that it can effectively capture micro-faults with a voltage drop as low as 30 mV within just a few seconds. Furthermore, the proposed method is inherently highly robust and is able to effectively detect false faults and hidden faults under varying current loads, which highlights the superiority of this method.
\end{abstract}


\markboth{}%
{}

\definecolor{limegreen}{rgb}{0.2, 0.8, 0.2}
\definecolor{forestgreen}{rgb}{0.13, 0.55, 0.13}
\definecolor{greenhtml}{rgb}{0.0, 0.5, 0.0}

\section{Introduction}

\begin{figure}[!t]\centering
	\includegraphics[width=8.5cm]{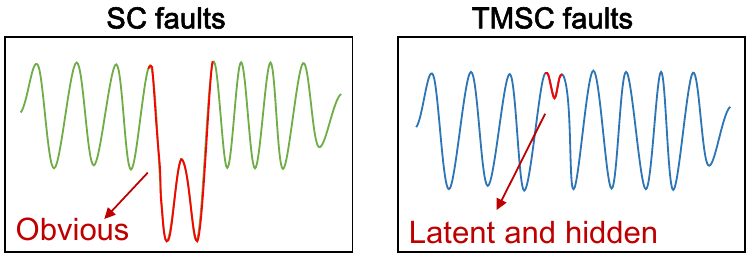}
	\caption{Voltage profile of TMSC faults.}\label{FIG_1}
\end{figure}

\IEEEPARstart{W}{ith}  the development of new green energy storage technologies, lithium ion batteries (LIBs) have become the key to the field of energy storage \cite{ref1}. Nevertheless, thermal runaway resulting from battery failure has become the most significant impediment to further development. Extensive re-search has determined that short circuit (SC) faults in batteries are the predominant cause of failure \cite{ref2}. Unlike the common SC faults, however, there is a special fault type exists in practical scenario, called transient micro short-circuit (TMSC) faults, as shown in Fig. \ref{FIG_1}, which is characterized by two features as follows:
\begin{itemize}
    \item Small drop in faulty voltage, typically 20 to 30 mV
    \item Short duration, typically a few sampling periods (e.g., 1 to 2 seconds for 1 Hz sampling).
\end{itemize}

Such latent faults are often highly concealed, particularly under dynamic discharging conditions. Moreover, considering the timing of sampling uploads, these faults should be identified at the management system level rather than being transmitted to the cloud platform for processing. Hence, it is vital to find an effective approach to capture them in time once such faults occur. 

Battery faults during operation may induce various abnormalities, including voltage jumps, thermal expansion, and gas generation \cite{ref3}. Voltage characteristics have been regarded as the crucial and significant feature for SC faults, driving widespread adoption of voltage-based diagnostics \cite{ref4}. Battery pack diagnostics typically leverage inter-cell voltage differences or clustering algorithms \cite{ref5}, yet such pack-dependent methods exhibit limitations in individual cell analysis under restricted data conditions. Voltage entropy metrics, measuring time-series disorder during faults \cite{ref6}, \cite{ref7}, provide an alternative with computational efficiency, though sensitivity degrades under minor voltage deviations or transient currents. Besides, equivalent circuit models-based approaches demonstrate enhanced adaptability in battery systems through various state observer technologies \cite{ref8}, \cite{ref9}, with validated practical viability. Finally, in battery fault diagnosis, machine learning techniques have demonstrated exceptional capabilities, including convolutional neural networks (CNN) \cite{ref10}, impedance-based deep neural networks (DNN) \cite{ref11}, and autoencoder (AE) frameworks \cite{ref12}, etc. Despite their high accuracy in battery energy storage system diagnostics \cite{ref13}, challenges persist in balancing computational complexity with real-world deployment requirements.

To address these challenges, this letter introduces an effective detection method for TMSC faults in LIBs. The proposed method uses continuous wavelet transform (CWT) for feature extraction of voltage and current signals of the batteries and detects micro faults by analyzing the con-sistency of the wavelet spectrum of both. The advantages of the proposed method are as follows: 

\begin{enumerate}[1)]
	\item  Simplicity, this method leverages signal processing techniques without relying on accurate battery model parameters.
	\item High sensitivity, it can precisely capture the highly covert faults, e.g., TMSC faults.
        \item High robustness, it can still effectively detect faults and avoid misdiagnosis under some extreme current load variations or abnormal disturbance.
\end{enumerate}

\begin{figure}[!t]\centering
	\includegraphics[width=8.5cm]{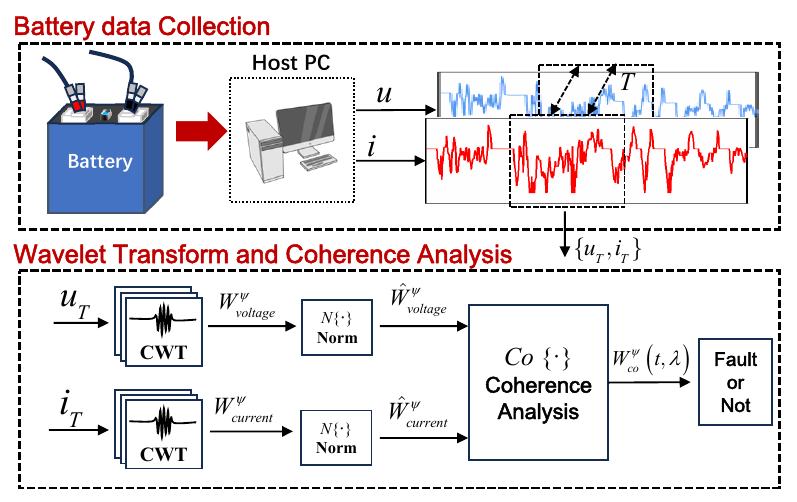}
	\caption{Process of fault detection.}\label{FIG_2}
\end{figure}

\section{Methodology of Fault Detection}

As shown in Fig. \ref{FIG_2}, the detection process is described as follows: firstly, the voltage and current of the battery are performed CWT to obtain the wavelet spectrum of the two, and then the wavelet spectrum is analyzed for coherence, finally, the fault is diagnosed based on the coherence of the two.

\subsection{Continuous Wavelet Transform}

CWT is a signal processing tool that allows time-frequency analysis of non-smooth signals. Consider a signal $x(t)$, and Fourier transform is defined by
\begin{align}
    X(\omega) = \int_{R} x(t) e^{-j\omega t} dt \label{eq1}
\end{align}
To add the time dimension, let us consider an admissible wavelet $\psi(t)$ satisfying $ 0 < \int_{0}^{+\infty}\omega^{-1}|\psi(\omega)|^2d\omega < \infty.$ For any time $t$ and scale $\lambda$, the CWT of origin $x(t)$ is defined by
\begin{align}
   W_x^{\psi}(t,\lambda)=\frac{1}{\sqrt{\lambda}}\int_R x(\tau)\psi^*\left(\frac{\tau - t}{\lambda}\right)d\tau
\end{align}
where $z^*$ represents the complex conjugate of $z$ and $\psi(t)$ is the mother wavelet. In addition, the CWT of $x(t)$ can also be realized in the frequency domain
\begin{align}
   W_x^{\psi}(t,\lambda)=\int_RX(\omega)\Psi^*(\lambda\omega)e^{j\omega t}d\omega
\end{align}
Here, this letter selects the Morlet wavelet as the mother wavelet, which optimizes the product of the time and frequency resolutions of the wavelet.
\begin{figure}[!t]\centering
	\includegraphics[width=8.5cm]{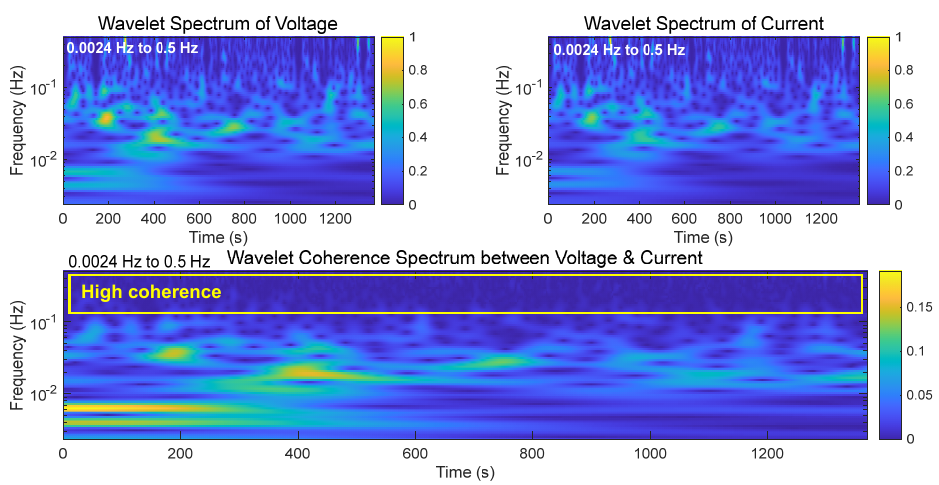}
	\caption{Wavelet spectrum and coherence analysis of a normal cell}\label{FIG_3}
\end{figure}
\begin{figure}[!t]\centering
	\includegraphics[width=8.5cm]{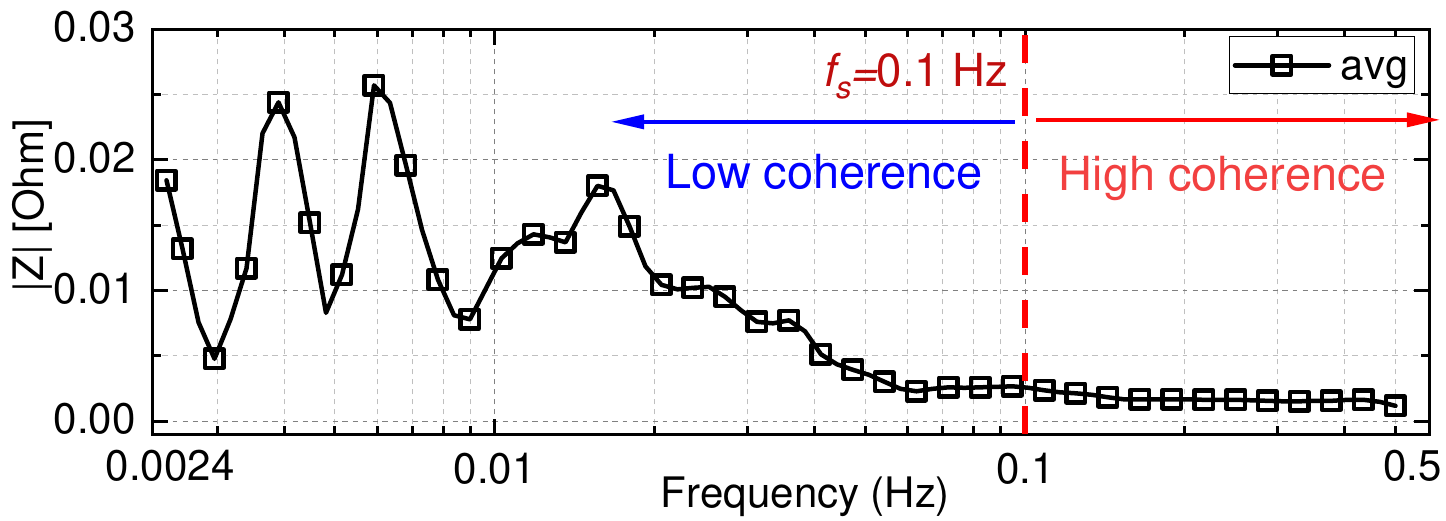}
	\caption{The average value of wavelet coherence spectrum between voltage and current within the period $T$}\label{FIG_4}
\end{figure}

\subsection{Wavelet Coherence Analysis}

In battery fault diagnosis tasks, voltage is usually regarded as an important feature and is widely used for analysis. However, the importance of current is often overlooked. In fact, based the waveforms of voltage and current, the two show a high degree of correlation in many cases, especially under complex dynamic operating conditions. Therefore, fault detection can be carried out by the coherence between the CWT spectrum of voltage and current. To reduce computational resources, the battery data {$u_T$, $i_T$} within the time window $T$ is selected for signal processing. In this study, we set $T$=1370 s, a standard length cycle for FUDS.

The CWT results of voltage and current{$u_T$, $i_T$} are denoted as $W_{voltage}^\psi(t,\lambda)$ and $W_{current}^\psi(t,\lambda)$ respectively. After CWT, the time-frequency distributions of these two physical quantities are two-dimensional matrices stored as complex values. In order to maintain subsequent coherence, it is necessary to normalize the complex matrix, which is expressed as follows:
\begin{align}
    \begin{cases}
     \hat{W}_{voltage}^{\psi}(t,\lambda) = \frac{W_{voltage}^{\psi}(t,\lambda) - \min\limits_{t,\lambda} | W_{voltage}^{\psi} |}{\max\limits_{t,\lambda}{| W_{voltage}^{\psi}|} - \min\limits_{t,\lambda} |W_{voltage}^{\psi} |}
    \\
     \hat{W}_{current}^{\psi}(t,\lambda) = \frac{W_{current}^{\psi}(t,\lambda) - \min\limits_{t,\lambda} | W_{current}^{\psi} |}{\max\limits_{t,\lambda}{| W_{current}^{\psi}|} - \min\limits_{t,\lambda} |W_{current}^{\psi} |}
    \end{cases}
\end{align}
where $\hat{W}$ represents the globally normalized matrix, and $|W|$ denotes the modulus of each element of the matrix, and $\max$ and $\min$ denote the maximum and minimum values of the matrix, respectively.

Finally, the CWT coherence matrix between voltage and current is expressed as:
\begin{align}
    W_{co}^\psi(t,\lambda)=||\hat{W}_{voltage}^{\psi}(t,\lambda)|-|\hat{W}_{current}^{\psi}(t,\lambda)||
\end{align}
Note that here each element of the coherence matrix $W_{CO}^\psi(t,\lambda)$ is a scalar. Fig. \ref{FIG_3} illustrated the coherence analysis of wavelet transform for the normal battery, and Fig. \ref{FIG_4} shows average value of wavelet coherence spectrum between voltage and current within the period T. It can be seen that at 1 Hz sampling frequency, a specific frequency $fs$ = 0.1 Hz can be selected as the demarcation between high and low coherence region. According to the above results, under normal conditions, the CWT coherence matrix of voltage and current signals naturally has high coherence in the high frequency region. This kind of TMSC faults is essentially an abnormal high-frequency signal (relative the voltage signal), so it can be detected them according to the coherence of CWT. If the coherence in the high-frequency region decreases at a certain time t, it indicates that there may be a fault. 

\section{Experimental Validation}
\subsection{Experiment}

\begin{figure}[!t]\centering
	\includegraphics[width=8.5cm]{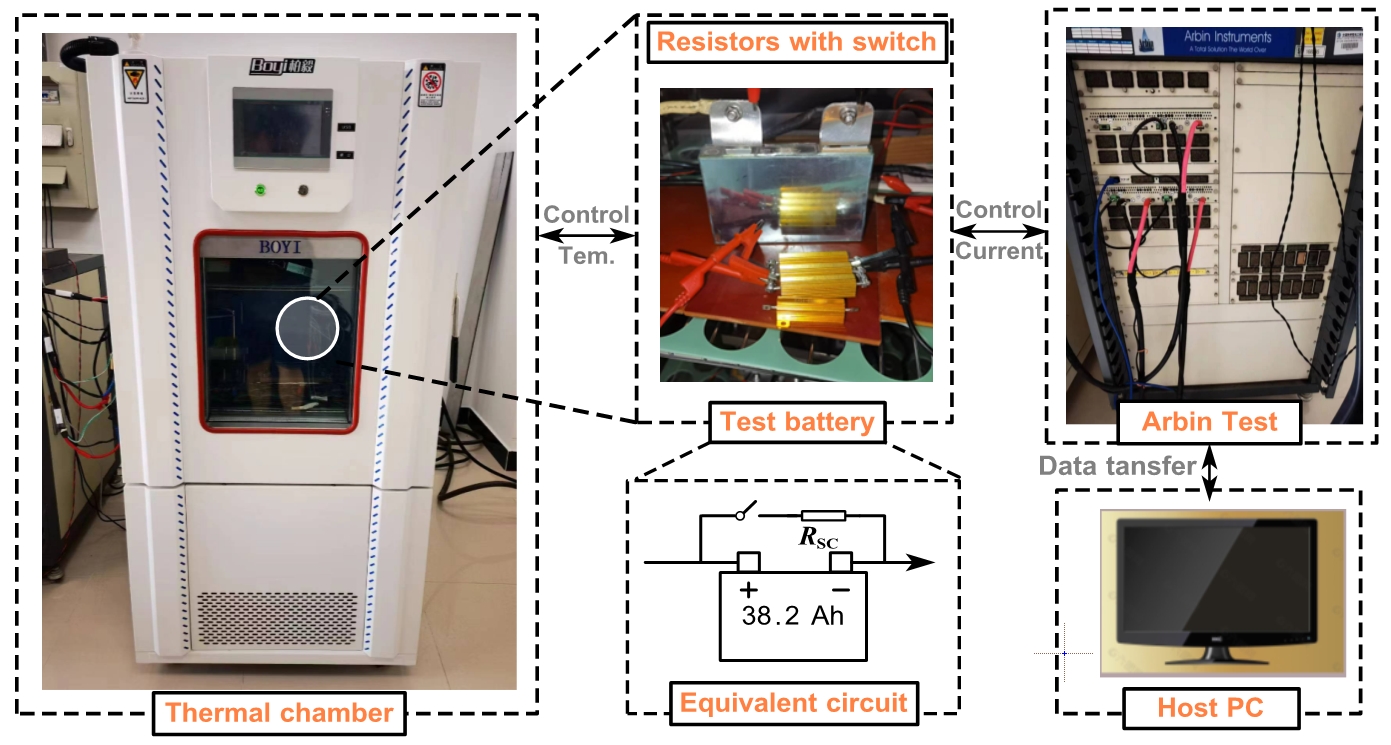}
	\caption{Fault experimental device schematic}\label{FIG_5}
\end{figure}

\begin{table}[!t]
    \renewcommand{\arraystretch}{1.8}
    \caption{Experiment Description}
    \centering
    \label{tab:experiment}
    \resizebox{\columnwidth}{!}{%
        \begin{tabular}{cllc}
            \hline\hline 
            \multicolumn{1}{c}{Number} & \multicolumn{1}{c}{Type} & \multicolumn{1}{c}{Description} & \multicolumn{1}{c}{Trigger Time} \\
            \hline
            \#1 to \#3 & TMSC Fault
            & \pbox{5cm}{Triggering micro faults by external \\ short circuit resistor \SI{500}{\milli\ohm}} 
            & \pbox{5cm}{\SIrange{501}{503}{s} \\ \SIrange{903}{906}{s} \\ \SIrange{1002}{1005}{s}} \\
            \hline
            \#4 & False Fault 
            & \pbox{5cm}{Triggering by applying pulse \\ discharging current \SI{60}{A}} 
            & \SIrange{621}{625}{s} \\
            \hline
            \#5 & Hidden Fault 
            & \pbox{5cm}{Simultaneous short circuit \SI{100}{\milli\ohm} \\ and pulse charging \SI{40}{A}} 
            & \SIrange{1021}{1024}{s} \\
            \hline\hline
        \end{tabular}%
    }
\end{table}

Fig. 5 shows the experimental device. In the experiment, we select a 38.2 Ah Ni\textsubscript{6}Co\textsubscript{2}Mn\textsubscript{2} lithium-ion battery as the test cell. The cell is discharged under the federal urban driving schedule (FUDS) condition, 1370 s a cycle. The Arbin 2000BT equipment controlled charging and discharging with a voltage sampling accuracy of 1 mV and a sampling period of 1 Hz. The external resistors trigger TMSC faults, manually connected to battery tabs. The experiment was conducted for two FUDS cycles with a total of five fault points. Table 1 demonstrates the specific steps. In the first fault cycle, three fault triggers were performed by simulating a micro-short circuit fault using a 1 $\Omega$ resistor, each lasting 2 to 3 s. In the second cycle: A pulse discharge with a higher current (60 A) was set at 621 s to simulate a load shock leading to a \textquotedblleft{}false\textquotedblright{} fault A SC fault by 100 m$\Omega$ resistor was triggered simultaneously with a 40A pulse charge at 1022 s to simulate a\textquotedblleft{}hidden\textquotedblright{} fault

\subsection{Fault Detection Results}

\begin{figure}[!t]\centering
	\includegraphics[width=8.5cm]{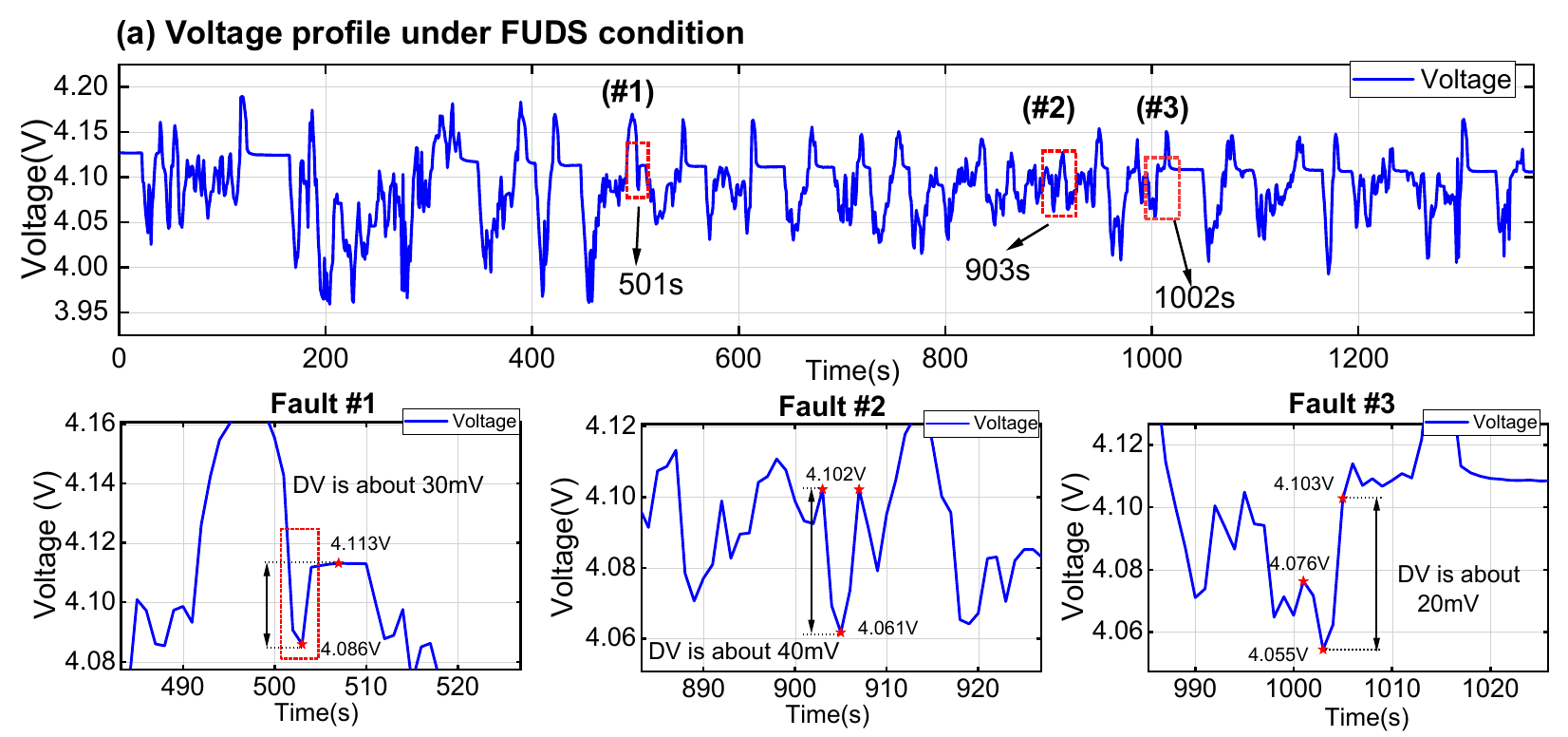}
	\caption{Voltage profile with TMSC faults under FUDS condition}\label{FIG_6}
\end{figure}
\begin{figure}[!t]\centering
	\includegraphics[width=8.5cm]{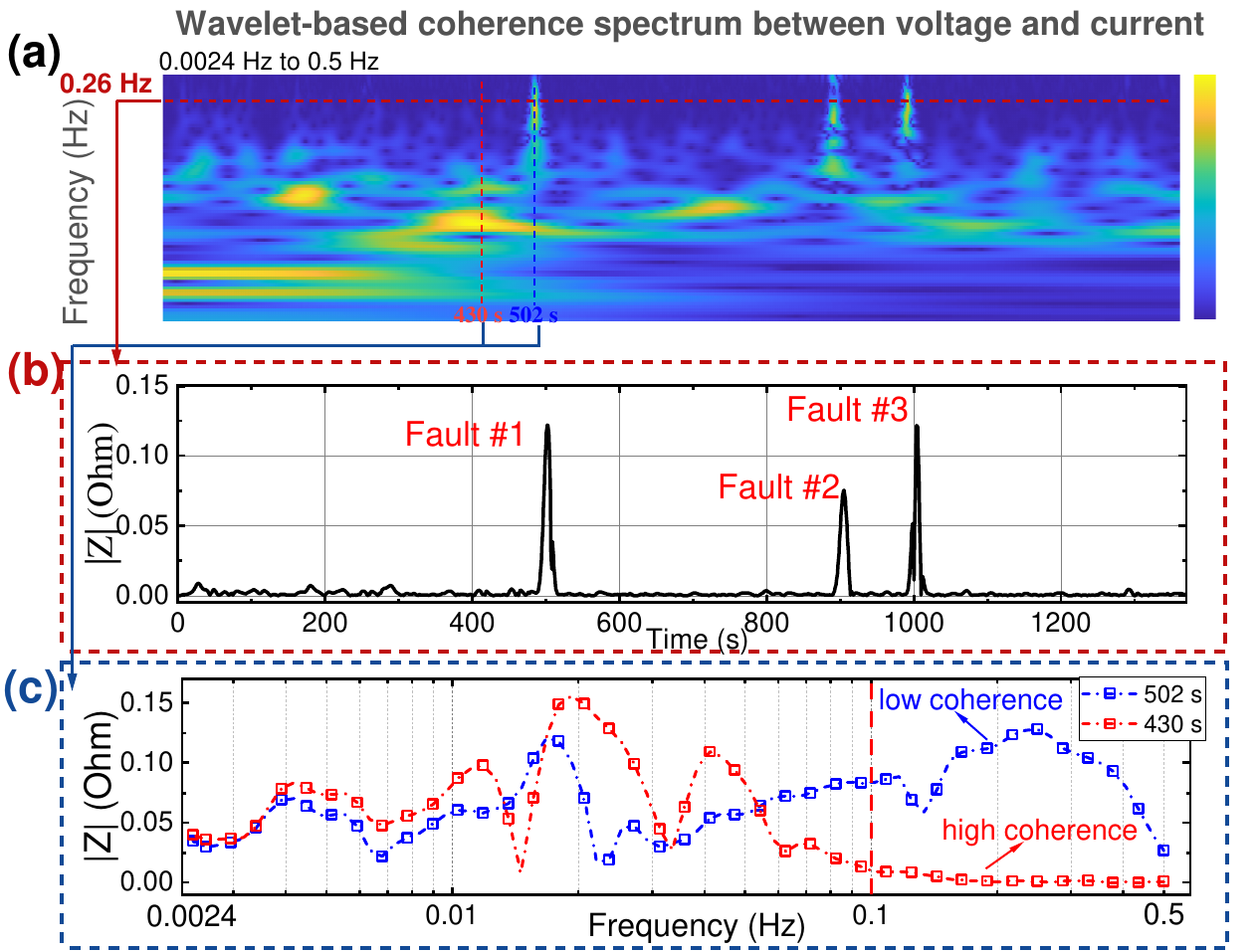}
	\caption{Detection results of TMSC faults. (a) Wavelet-based coherence spectrum. (b) Wavelet spectrum results at 0.26 Hz. (c) Wavelet spectrum results at 430 s and 502 s}\label{FIG_7}
\end{figure}

Fig. \ref{FIG_6} presents the voltage profile under three covert TMSC faults. Notably, under dynamic conditions, the ab-normal voltage fluctuations (i.e., red box) induced by TMSC faults are concealed within the normal voltage pulsations, rendering them exceptionally challenging to distinguish. It can be seen in Fig. 6 that the voltage drop (DV) caused by TMSC fault is about 20 to 30 mV. The detection results of TMSC faults are illustrated in Fig. \ref{FIG_7} and in Fig. \ref{FIG_7}(a), three distinct high-energy regions (0.1 to 0.5 Hz) are observed in the spectral domain, exhibiting marked contrast to the low-energy dark background. These observations confirm three low-coherence regions between voltage and current signals, directly attributable to the abnormal voltage variations caused by TMSC faults. Fig. \ref{FIG_7}(b) further demonstrates three distinct energy peaks at 0.26 Hz, quantitatively confirming the presence of three anomalies. To validate the consistency of this analysis, wavelet spectrum results at 430 s and 502 s were selected, which align with the aforementioned findings. This confirms that covert faults can be reliably detected through coherence analysis within the high-frequency spectrum.

This method can detect such faults very effectively, but it is worth noting that because the sampling period is 1 Hz, the maximum spectrum frequency according to Shannon sampling theorem is 0.5 Hz, which is 0.0024 to 0.5 Hz in this study. If the sampling frequency changes, the relative high frequency region also changes.

\subsection{Robustness Validation}
\begin{figure}[!t]\centering
	\includegraphics[width=8.5cm]{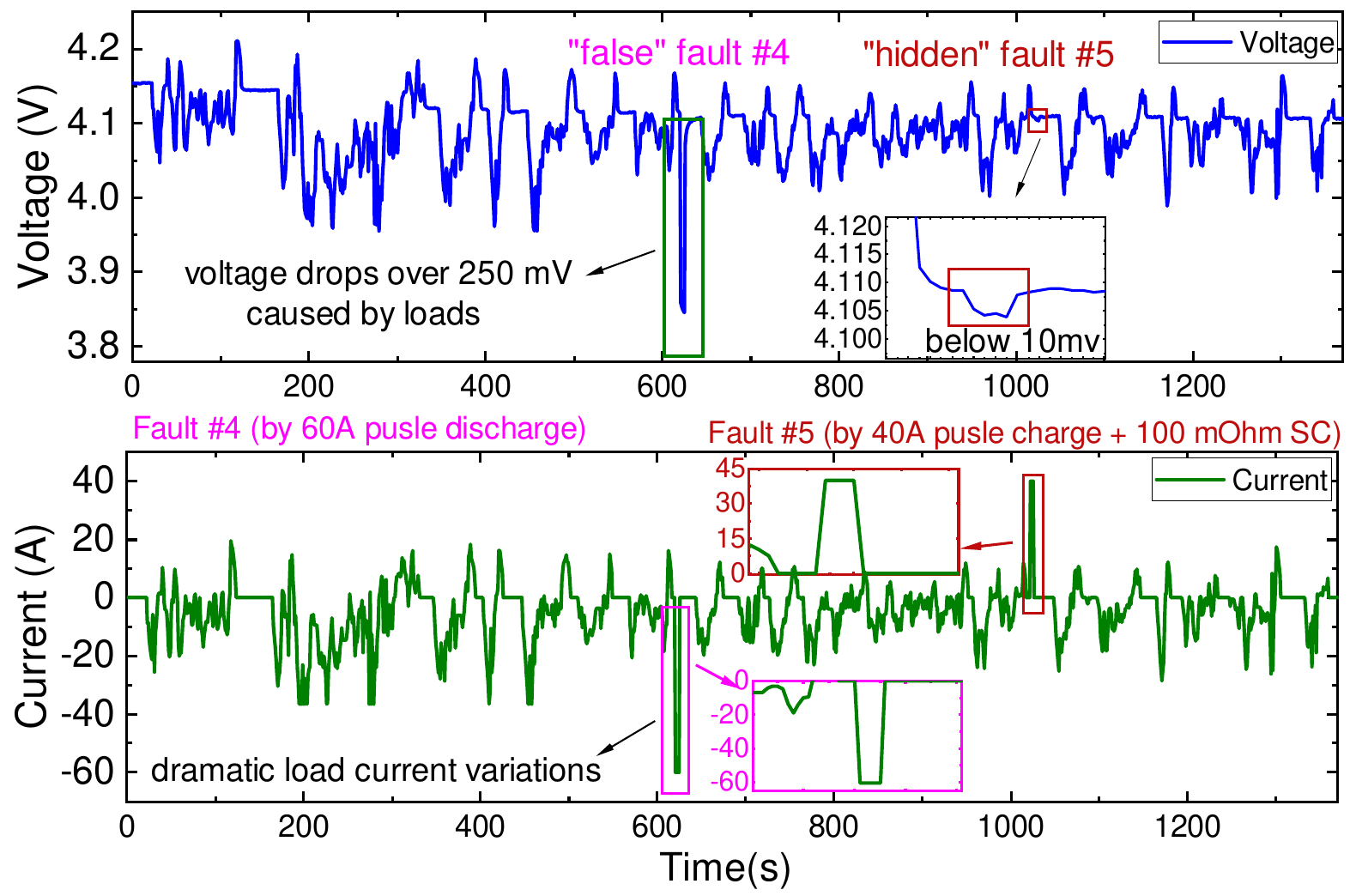}
	\caption{Voltage and current profile with \textquotedblleft{}false\textquotedblright{} and \textquotedblleft{}hidden\textquotedblright{} faults}\label{FIG_8}
\end{figure}

Let us recall the core idea of the proposed method: detecting anomalies based on the coherence of voltage and current at a specific frequency. This shows that this method is still robust to some specific situations and disturbances in practical applications. Two special scenarios are shown in Fig. \ref{FIG_8}, one with abnormal voltage changes due to abnormal current load variations (\textquotedblleft{}false fault\textquotedblright{}) or the other with short-circuit faults and load variations occurring at the same time (\textquotedblleft{}hidden fault\textquotedblright{}).

\begin{figure}[!t]\centering
	\includegraphics[width=8.5cm]{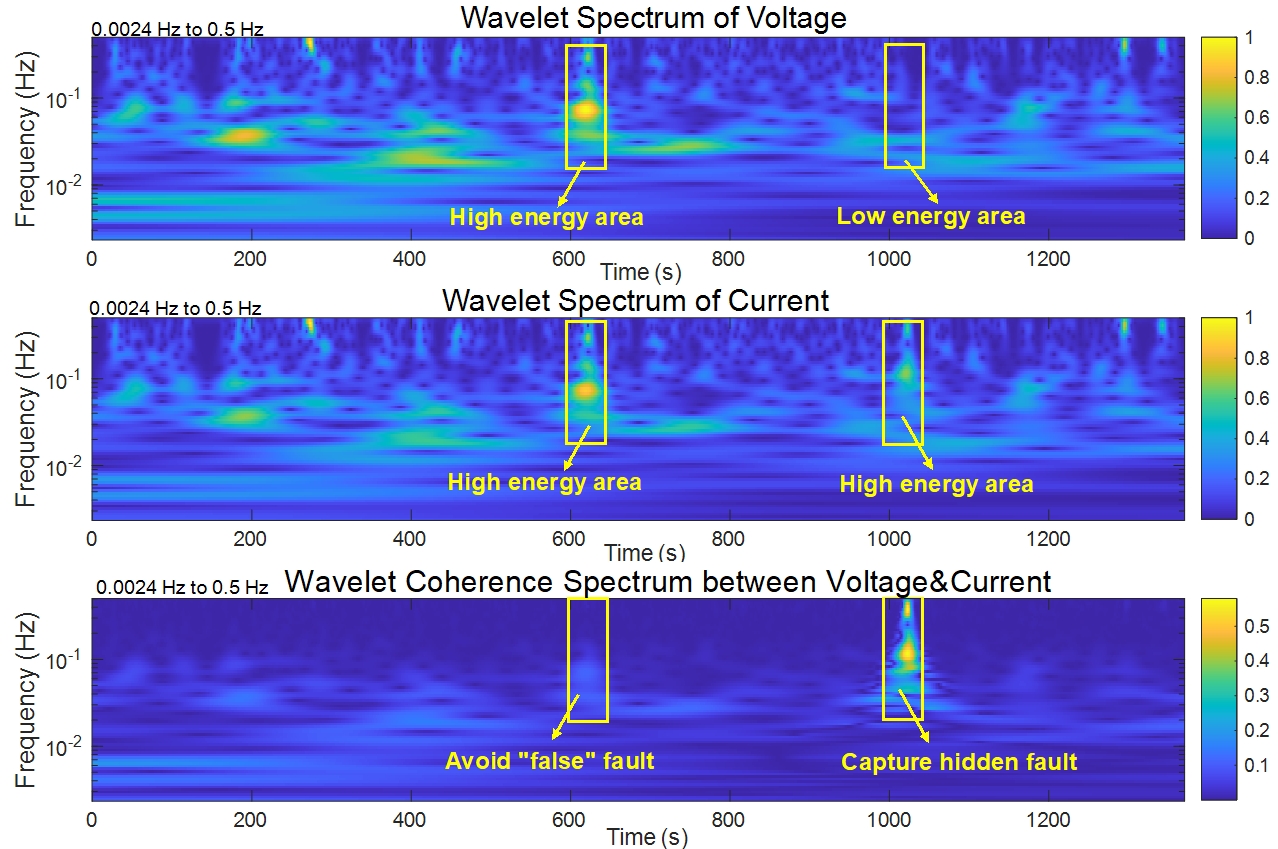}
	\caption{Wavelet spectrum in special scenarios}\label{FIG_9}
\end{figure}

Fig. \ref{FIG_9} shows the wavelet spectrum in special scenarios, including a \textquotedblleft{}false\textquotedblright{} fault and a hidden fault. It can be seen that for fault \#4, although the voltage fluctuates abnormally resulting in highlighted energy regions in the wavelet spectrum, the current spectrum is also keep high coherence, so the final coherence spectrum still does not have high-energy regions in the high frequency, indicating that misdiagnosis for \textquotedblleft{}false\textquotedblright{} faults can be avoided effectively. Similarly, for \textquotedblleft{}hidden\textquotedblright{} fault, although the voltage does not show any significant abnormal voltage fluctuation (<10 mV), the inconsistency with the current results in high-energy regions at the high-frequency in the coherence spectrum. This shows that this method can effectively deal with a variety of situations and has a high level of anti-interference capability and robustness.

\section{Discussion}
The signal processing-based method proposed in this paper can be very effective in capturing those hidden faults that are difficult to find in practice, and the effectiveness is verified by experiments. However, there are still two key points worth discussing, as follows: 1) Considering the terminal computational resources and commercial cost, discrete wavelet variation (DWT) can be chosen as the algorithm in BMS; 2) Actual battery operation is complex, and validation of the method requires more hidden and micro-failure types which need to further investigation. We will continue to address these issues in future work, and the methodology proposed in this study offers a new perspective for analyzing such micro-faults.

\section{Conclusion}

This study proposes a method based on signal processing for accurately detecting TMSC faults of LIBs. By analyzing the coherence of voltage and current signals in the wavelet spectrum at specific frequencies, micro-scale anomalies can be rapidly identified. Experimental results show that the method is capable of detecting voltage drops as low as 30 mV within seconds, with strong robustness in distinguishing true faults from false or hidden ones under varying load conditions. This research not only enhances the safety of battery systems but also provides a new perspective for algorithm commercialization and practical applications.


\bibliographystyle{IEEEtranTIE}

\end{document}